\documentclass[times, 10pt,twocolumn]{article} 
\usepackage{latex8}
\usepackage{times}
\usepackage{graphicx}
\usepackage[hyphens]{url}
\usepackage{caption}
\usepackage{subcaption}
\usepackage{algpseudocode}
\usepackage{amsmath}
\usepackage{varwidth}
\usepackage{enumitem}

\usepackage{mathtools}
\usepackage{csquotes}
\usepackage{float}
\usepackage{amsmath}
\usepackage{graphicx}
\usepackage{caption}
\usepackage{newfloat}
\usepackage{amssymb}
\usepackage{subcaption}
\usepackage{multirow}
\usepackage{alltt}
\usepackage{upquote}
\usepackage[table]{xcolor}
\usepackage{epstopdf}
\usepackage{tabularx}
\usepackage{booktabs}
\usepackage{url}
\usepackage{enumerate}

\begin{document}

\title{A Covert Data Transport Protocol}

%\author{Yu Fu\\
%The Holcombe Department of Electrical and Computer Engineering\\ Clemson University \\ fu2@g.clemson.edu\\
%\and
%Zhe Jia\\
%Department of Physics and Astronomy\\
%Clemson University \\
%zhej@g.clemson.edu\\
%\and
%Lu Yu, Xingsi Zhong, Richard Brooks\\
%The Holcombe Department of Electrical and Computer Engineering\\
%Clemson University \\
%lyu,xingsiz, rrb@g.clemson.edu\\
%}

\author{Yu Fu$^{*}$, Zhe Jia$^{\dagger}$, Lu Yu$^{*}$, Xingsi Zhong$^{*}$, and Richard Brooks$^{*}$ \\${}$\\
$^{*}$ The Holcombe Department of Electrical and Computer Engineering \\ \{fu2, lyu, xingsiz, rrb\}@g.clemson.edu\\${}$\\
$^{\dagger}$Department of Physics and Astronomy \\ zhej@g.clemson.edu \\
Clemson University, Clemson, SC, 29634, USA
}

\maketitle
\thispagestyle{empty}

\begin{abstract}
Both enterprise and national firewalls filter network connections. For data forensics and botnet removal applications, it is important to establish the information source. In this paper, we describe a data transport layer which allows a client to transfer encrypted data that provides no discernible information regarding the data source. We use a domain generation algorithm (DGA) to encode AES encrypted data into domain names that current tools are unable to reliably differentiate from valid domain names. The domain names are registered using (free) dynamic DNS services. The data transmission format is not vulnerable to Deep Packet Inspection (DPI). 
\end{abstract}

\section{Introduction}\label{intro}
Protocol obfuscation is widely used for evading censorship and surveillance, and hiding criminal activity. Most firewalls use DPI to analyze network packets and filter out sensitive information. But if the source protocol is obfuscated or transformed into a different protocol, detection techniques that worked well with the source protocol will either detect nothing sensitive, or detect something which is far from the real information. When encrypted connections draw attention or are blocked \cite{iran}, protocol obfuscation is a solution for covert communication. For example, consider a botnet mothership trying to diffuse instructions to its infected zombies, obfuscating encrypted data streams is very useful.

In general, protocol obfuscation can be divided into two categories: (1) protocol mimicry, and (2) protocol tunnelling. Protocol mimicry is to make protocol A look like protocol B, by tampering some features (packet syntax and statistical features) of protocol A. Protocol tunnelling is to take the contents of packets from protocol A and put them into the payload of protocol B. In this way, protocol B is a carrier and masquerade of protocol A. As Houmansadr \cite{houmansadr2013parrot} pointed out, most protocol mimicry fails to be completely unobservable even without the attacker resorting to correlating multiple network flows or performing sophisticated traffic analysis. He concluded that mimicking the protocol in its entirety, including its reaction to errors, typical traffic and artifacts, is difficult. Protocol tunnelling, on the other hand, is easy to implement and there are many tools available. However, it is vulnerable to statistical analysis \cite{ born2010detecting, farnham2013detecting}.

The covert data transport protocol we present transforms arbitrary network traffic into legitimate DNS traffic. The server encodes the message into a list of domain names and register them to a randomly chosen IP address. The client does a reverse-DNS lookup on the IP address and decodes the domain names to retrieve the message. Different from DNS tunneling, this doesn't use uncommon record types (TXT records) or carry suspiciously large volume of traffic as DNS payloads. On the contrary, the resulting traffic will be normal DNS lookup/reverse-lookup traffic, which will not attract attention. The data transmission is not vulnerable to DPI. 

The structure of this paper is as follows. In Section \ref{related}, protocol obfuscation techniques are discussed. Section \ref{background} provides the background for this work. Section \ref{method} proposes the transport protocol. Section \ref{conclusion} concludes the paper and suggests future work.

\section{Related Work}\label{related}

\subsection{Protocol obfuscation techniques}

As one of the most famous anonymization tools, the Onion Router (Tor) \cite{tor} provides an infrastructure for anonymous communication over a public network. Obfsproxy \cite{obfsproxy} circumvents censorship by camouflaging Tor traffic between client and bridge nodes. It supports multiple protocols, called pluggable transports, which specify how traffic is transformed. ScrambleSuit \cite{winter2013scramblesuit} is a network protocol to obfuscate the transported application data to defend against active probing and protocol fingerprinting. SkypeMorph \cite{mohajeri2012skypemorph} is a Tor pluggable transport to reshape Tor packets to resemble Skype calls. StegoTorus \cite{weinberg2012stegotorus} first uses chopping to change packet sizes and timing information, and then uses steganography to disguise Tor traffic as a message in an innocuous cover protocol, such as HTTP. Format-Transforming Encryption (FTE) \cite{dyer2013protocol} evades regular-based deep packet inspection (DPI) technologies by transforming Tor traffic into a predefined format. Dust \cite{wiley2011dust} provides blocking resistance against the most common packet filtering techniques. 

Tor has been used by botnets to help hide their command and control (C\&C) nodes (motherships) \cite{tor1, casenove2014botnet, tor3}. In August 2013, Tor had seen a rapid spike of clients, which turned out to be click-fraud botnet running its C\&C as a Tor Hidden Service \cite{tor2, hopper2013protecting}.

Other protocol obfuscation techniques including randomized flushing of data streams and random padding \cite{hjelmvik2010breaking} are widely used by P2P software, for example, Skype, BitTorrent, and eMule. Decoy routing \cite{karlin2011decoy} makes it possible for a client to connect to any unblocked host/service as a middle point and finally connect to a blocked destination without cooperation with the host. Similar to protocol tunneling, Houmansadr \cite{houmansadr2013parrot} suggests running the actual protocol to embed data rather than mimicking the protocol. It is not certain that this would be practical, since the resulting combined protocol would contain side channels that could be modeled using the cross-product of two Markov models \cite{bhanu2010timing}. 

Obfuscation techniques in the last paragraph hide the protocol being used, but do not hide the address of the communication partner. Tor hides the address of the communication partners, and may hide the fact that Tor is being used. Mainly, Tor hides the communication partners by adding two extra network hops and three encryptions. Many countries block Tor used by stopping access to all Tor entrance IP addresses. The approach we present hides the identity of the partner, hides the protocol being used, and produces common DNS traffic.

\subsection{DNS-based steganographic channels}
Since our paper focuses on transforming arbitrary network traffic into DNS traffic, it is necessary to look at DNS-based steganographic channels. The Feederbot malware \cite{dietrich2011botnets} implements a transmission scheme by encoding messages into DNS TXT records, with which the bots query DNS servers for C\&C commands. PlugX \cite{cole2003hiding}, a remote access tool providing remote control and surveillance capabilities, uses DNS as a carrier protocol for C\&C communication. Thyer \cite{thyer2008covert} and Altalhi \cite{altalhi2011dns} use DNS 16-bit identification (ID) in the IP header as a covert channel to transmit data, which only encodes two characters of message in each packet. Ngadi \cite{ngadi2011indirect} uses DNS packet length covert channel, which is suitable for short message transfer. Most DNS-based steganographic channels suffer from low throughput or side-channel analysis. For example, DNS TXT-based steganographic channels have different payload length (probably longer) than normal DNS traffic, which are subject to side-channel analysis in packet length. DNS packet length-based steganographic channels have different packet length distribution than normal DNS traffic, which are subject to side-channel analysis in the entropy of the packet length distribution. Since our approach produces real DNS responses, it generates real DNS traffic, which is not subject to side-channel analysis.

\section{Background}\label{background}
\subsection{Hidden Markov Models (HMMs)}
A Markov model is a tuple $G = (S, T, P )$ where $S$ is a set of states of a model, $T$ is a set of directed transitions between the states, and $P=\left\{ p(s_i,s_j)\right\} $ is a probability matrix associated with transitions from state $s_i$ to $s_j$ such that:
\begin{equation}
\sum\nolimits_{s_j \in S} p(s_i,s_j) = 1, \forall s_i \in S
\end{equation}

A Markov model satisfies the Markov property, where the next state only depends on the current state. An HMM is a Markov model with unobservable states. A standard HMM \cite{eddy1996hidden, rabiner1989tutorial} has two sets of random processes: one for state transition and the other for symbol outputs. A deterministic HMM \cite{lu2013normalized, lu2012network, schwier2009pattern} is used in this paper, which only has one random processes for state transitions. Different output symbols are associated with transitions with different probability. This representations is equivalent to the standard HMM \cite{vanluyten2008equivalence}.

HMMs are widely used for pattern recognition and detection. Chen \cite{lu2012network} used HMMs to detect Zeus botnet zombies and achieved 94.7\% true positive rate (tpr) and 0.7\% false positive rate (fpr). Inter-packet delay of Zeus packets were collected and represented by an HMM, which was inferred with the zero-knowledge HMM inference algorithm \cite{yu2013inferring}. Zhong \cite{zhong2015side} conducted a side-channel attack on Phasor Measurement Units (PMUs) in an electric power grid and used HMMs to differentiate data source in an encrypted tunnel.

\subsection{Domain Generation Algorithms (DGAs)}
Botnets are groups of compromised computers that botmasters use to launch attacks over the Internet. In modern botnets, fast-flux is used to change the mapping between IP addresses and DNS names of the C\&C server periodically to avoid detection. DGAs are used to generate large numbers of DNS names as the rendezvous for the covert channel between bots and the botmaster. DGAs are widely used by botnets including Zeus \cite{andriesse2013highly}, Conficker \cite{shin2012large}, Kraken\cite{amini2008kraken}, Srizbi \cite{wolf2008technical}, Torpig \cite{stone2009your} etc.

Our previous work developed an HMM-based DGA, which evaded the DGA detection metrics (Kullback-Leibler distance, Edit distance, and Jaccard Index) in the literature \cite{yadav2012detecting} and two cutting-edge DGA-detection systems (BotDigger \cite{zhang2016botdigger} and Pleiades \cite{antonakakis2012throw}). In our previous HMM-based DGA work, we inferred an HMM that represents the linguistic features of legitimate domain names collected from the entire IPv4 space \cite{sonar}. Domain names generated by the HMM will be similar to, but not conflicting with legitimate domain names statistically. The inferred HMM will be used in this paper.

\subsection{Format-Transforming Encryption (FTE)}
FTE \cite{dyer2013protocol} extends conventional symmetric encryption by formatting the ciphertext. Arbitrary application-layer network traffic can be transformed into a target protocol using FTE. It is used to evade regular expression based protocol identification in Internet censorship and surveillance, because the original protocols are obfuscated. Zhong \cite{zhong2015stealthy} improved FTE to mimic arbitrary network protocols without learning their regular expressions. This work is similar to FTE, because we transform data into domain names, which also obfuscates the original protocol with DNS protocol. 

\section{Proposed Data Transport Protocol}\label{method}

\begin{figure}
\begin{center}
\includegraphics[width=.45\textwidth]{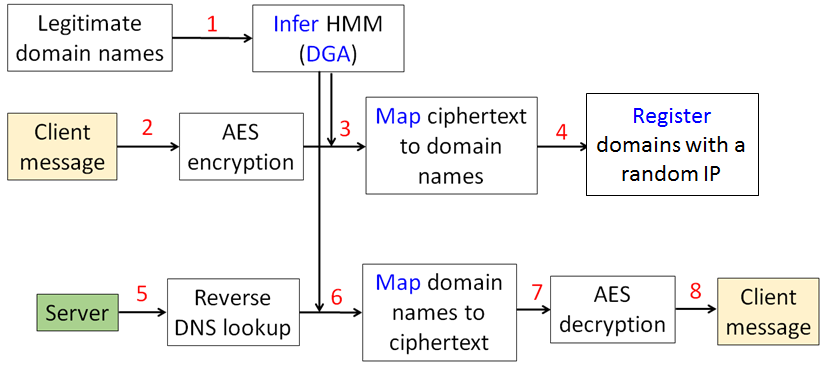}
\caption{Flow chart of the covert data transport protocol}
\label{flow-chart}
\end{center}
\end{figure}

The covert data transport protocol is based on two-way communications between a client and a server. Figure \ref{flow-chart} shows that the flow chart of the protocol. Before communication starts, the client and server will share the following information out of band: (1) AES key, (2) pseudo-random number generator (PRG), (3) PRG seed, and (4) an HMM that represents the statistical model of legitimate IPv4 domain names (step 1) and start state.

The HMM guarantees that domain names generated by the HMM will have the same statistical features as, but not conflict with, legitimate domain names. The steps of communication from the client to the server are:
\begin{itemize}[noitemsep,nolistsep]
\item The client prepares the message (`client message') to send to the server.
\item The message is AES encrypted into ciphertext (step 2).
\item The client maps ciphertext to domain names with the HMM-encoding algorithm described below (step 3).
\item The client registers the generated domain names with the pseudo-randomly chosen IP address (step 4).
\item The server reverse DNS-lookups the chosen IP address and retrieves the domain names (step 5).
\item The server maps domain names back to ciphertext (step 6).
\item Ciphertext is AES decrypted into client message (step 7 and 8).
\end{itemize}

\subsection{HMM-encoding/decoding algorithm}
HMM-encoding maps data strings (ciphertext in our work) to domain names. Since the HMM is a probabilistic regular grammar with transitions associated with different probability, an intuitive idea is to find a path in HMM on one side, which can be recovered on the other side. This requires both sides to choose a start state and the same encoding/decoding algorithm. We choose the state with the largest asymptotic probability as the starting state because it is the state occurring with the largest probability as time goes to infinity.

To encode the message into the path of HMM, we round transition probabilities to the closest $\frac{1}{2^n}$, where $n$ is an integer. Rounding keeps the statistical features of legitimate domain names, while finding a way to encode and decode. Since the original probabilities of all transitions going out of a state sum up to 1, it is possible to round them to the closest $\frac{1}{2^n}$. The set of transition probabilities associated with state $S_i$ is denoted as $\left\{ P_{i1}, P_{i2}, ... , P_{ik}\right\}$, where $k$ is the total number of transitions leaving state $S_i$ and $\sum_{n=1}^{k} P_{in} = 1$. The set of rounded transition probabilities associated with state $S_i$ is denoted as $\left\{ RP_{i1}, RP_{i2}, ... , RP_{ik}\right\}$, where $k$ is the total number of transitions leaving state $S_i$ and $\sum_{n=1}^{k} RP_{in} = 1$.

With the rounded transition probabilities of state $S_i$, we associate a binary representation for each transition. This allows us to encode binary data (converted from the original `client message') into the HMM. Figure \ref{bin} shows the binarization algorithm. The input is the rounded probability ($\left\{ RP_{im} \right\}$, $m = 1,...,k$), and the output is the binary representation ($\left\{ BIN\_RP_{im}\right\}$) for the corresponding transitions. Note that any string can be encoded with the binary representations.

\begin{figure}
\begin{algorithmic}[1]
\Procedure{binarization}{$\left\{ RP_{im} \right\}$}
\State order $\left\{ RP_{im} \right\}$ from the largest to the smallest into $\left\{ ORP_{i1}, ORP_{i2}, ... , ORP_{ik}\right\}$
\State start with the largest value $ORP_{i1}$, and $BIN\_{ORP_{i1}}$ = $\underbrace{0...0}_{\log_2 \frac{1}{ORP_{i1}}}$
%$BIN\_{ORP_{i1}}$ = $\log_2 \frac{1}{V_{i1}}$ of $0$s 
%\State $\underbrace{0...0}_{5}$
\State initiate j = 2, last\_input = $ORP_{i1}$, last\_output = $BIN\_{ORP_{i1}}$, and $num = \log_2 \frac{1}{ORP_{i1}}$

\While{j $\leq$ k} \Comment{not all values are assigned}

\If {$ORP_{ij}$ == last\_input:}
\State $BIN\_{ORP_{ij}}$ = last\_output + 1 

\Else
\State calculate $diff$ =  $\log_2 \frac{1}{ORP_{ij}}$ - num
\State $BIN\_{ORP_{ij}}$ = (last\_output + 1)*$2^{diff}$
\State update $num = \log_2 \frac{1}{ORP_{ij}}$

\EndIf 
\State update last\_input = $ORP_{ij}$
\State update last\_output = (last\_output + 1)*$2^{diff}$
\State j = j + 1

\EndWhile\label{euclidendwhile}
\State From the mapping between $\left\{ {RP_{ij}} \right\}$ to $\left\{ {ORP_{ij}} \right\}$, obtain $\left\{ BIN\_{RP_{ij}} \right\}$ from $\left\{ BIN\_{ORP_{ij}} \right\}$
\State \textbf{return} $\left\{ BIN\_{RP_{ij}} \right\}$
\EndProcedure
\end{algorithmic}
\caption{Proability binarization algorithm}\label{bin}
\end{figure}

After binarization, each transition of the HMM has a binary representation. Given the binary data, there is a unique path going through the HMM that encodes the binary data string. However, there are two technical difficulties:
\begin{itemize}[noitemsep,nolistsep]
\item The output symbols in HMM includes `a-z', `0-9', `-' and '$\textvisiblespace$' (space). Normally, domain names generated from the HMM are separated with space (for example `ab$\textvisiblespace$cd$\textvisiblespace$ef'. But the receiver will not be able to put space in the correct spot (only put space after `ab' and `cd', but not `ef').
\item With receiving the domain names (`ab', `cd' and `ef'), the receiver doesn't know the order.
\end{itemize}

\begin{figure}
\begin{center}
\includegraphics[width=.2\textwidth]{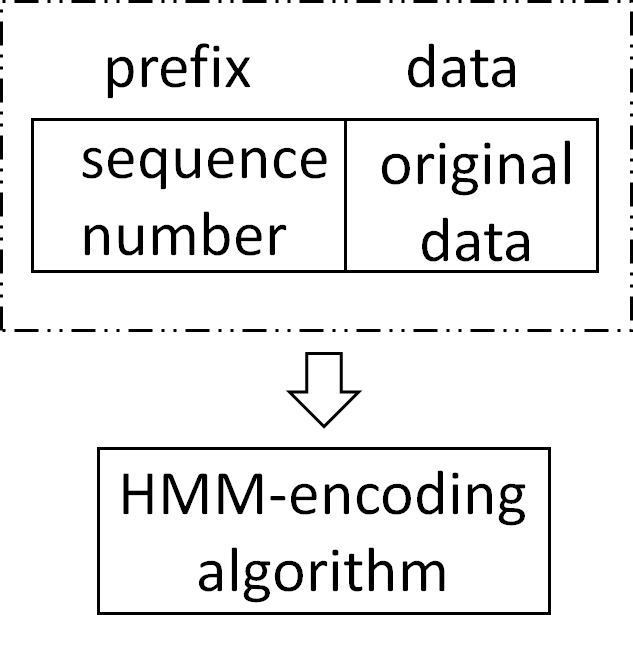}
\caption{The new data structure}
\label{new-ds}
\end{center}
\end{figure}

%The solution is to add the prefix and suffix to the original data. The new data structure is in Figure \ref{new-ds}. The prefix is randomly selected from a pool of characters that does not generate space output, which guarantees at least one character from the original data can be encoded. The suffix contains the sequence number, which guarantees the correct order of the message on the receiver side.

The solution is to add the prefix to the original data. The new data structure is in Figure \ref{new-ds}. The prefix is the sequence number, which guarantees the correct order of the message on the receiver side.

After the client message is converted into multiple domain names, we register them as sub-domain names with the server IP address using dynamic DNS service (https://freedns.afraid.org/ for example).

To decode the domain names into client message, the server will reverse-lookup a randomly chosen IP address and retrieve a list of sub-domain names. After doing the inverse of the encoding, several message pieces are recovered. Using the sequence number, the original client message is obtained.

Since both client and server will register the same randomly chosen IP address to make the system work, we have to consider:
\begin{itemize}[noitemsep,nolistsep]
\item The probability of choosing a collided domain name is almost zero if using IPv6 domain names. The IPv6 address space range contains $2^{128}$ entries, which is more than one entry for every three atoms in the universe. If an address is chosen at random, the chance of a collision is essentially zero. If the entire IPv4 space were to choose IPv6 addresses at random, the probability of at least one collision is approximately 0.0155.
\item If collision happens, the system will work fine. Because the HMM decoding won't work on the existing domain names. By simply ignoring the domain names that fail to pass HMM decoding, we can separate the desired domain names from the existing ones.
\end{itemize}

Besides the two-way communication, the idea can also be used in botnets where the botnet C\&C encodes the message into domain names, and the bot retrieves the message. This will make the botnet traffic look like pure DNS traffic.

\begin{figure}
\begin{center}
\includegraphics[width=.3\textwidth]{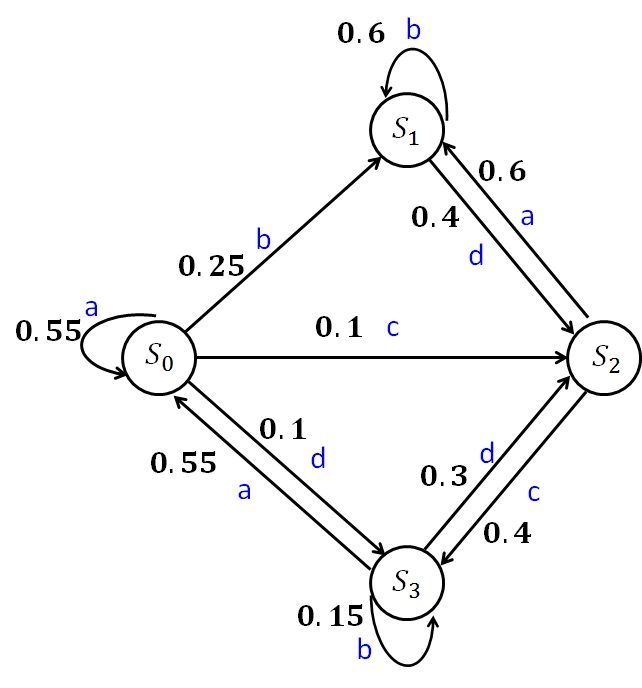}
\caption{An example HMM for illustration}
\label{ex-hmm}
\end{center}
\end{figure}

\begin{figure}
\begin{center}
\includegraphics[width=.3\textwidth]{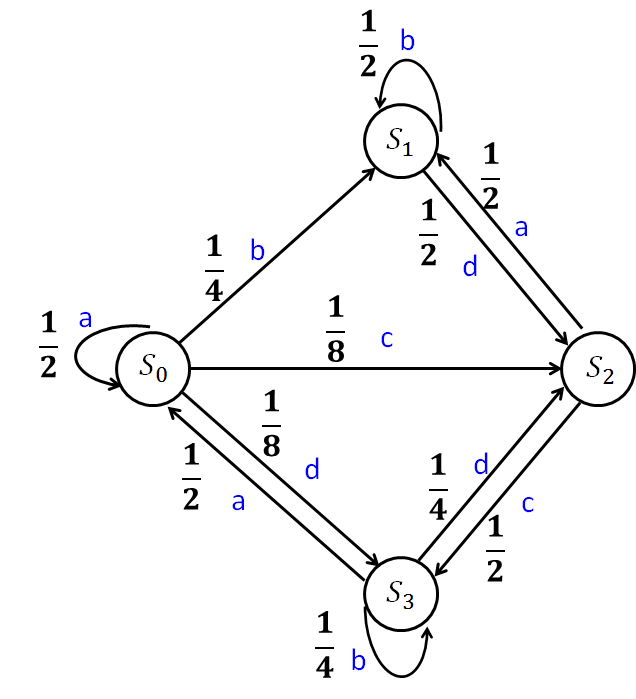}
\caption{HMM after probability rounding algorithm}
\label{round-fig}
\end{center}
\end{figure}

\begin{figure}
\begin{center}
\includegraphics[width=.3\textwidth]{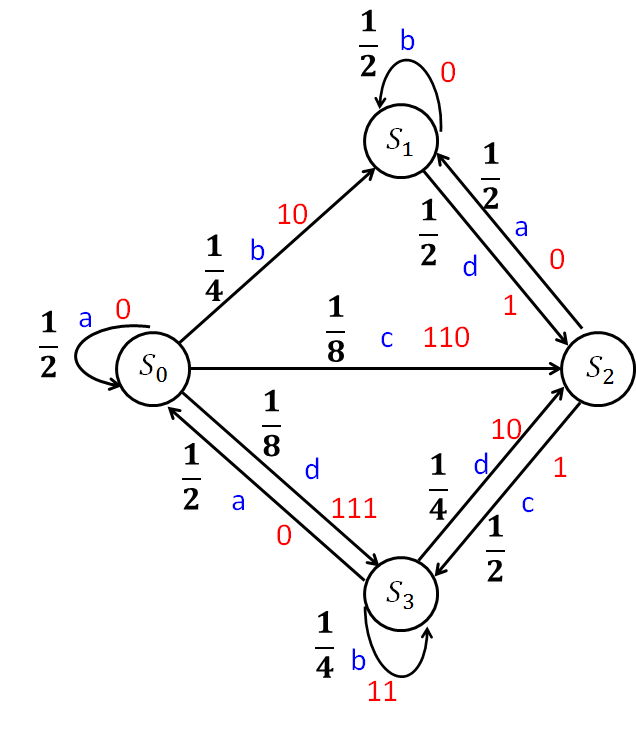}
\caption{HMM after binarization algorithm}
\label{bin-fig}
\end{center}
\end{figure}

\begin{figure}
\begin{center}
\includegraphics[width=.2\textwidth]{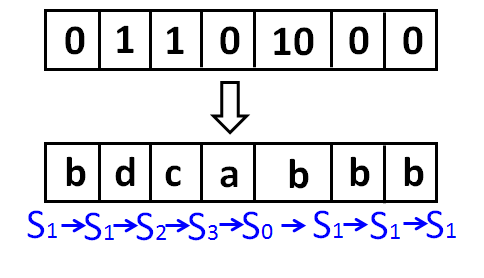}
\caption{An example mapping}
\label{map}
\end{center}
\end{figure}

\begin{table}[H]
\centering
\small
\caption{Performance of HMM encoding/decoding}
\label{perf}
%\begin{tabular}{|c|c|c|c|c|}
\begin{tabular}{|m{.09\columnwidth}|m{.09\columnwidth}|m{.13\columnwidth}|m{.13\columnwidth}|m{.32\columnwidth}|@{}l@{}}
\hline
\begin{tabular}[c]{@{}c@{}}Input\\ String\\ Length\end{tabular} & \begin{tabular}[c]{@{}c@{}}AESed\\ String\\ Length\end{tabular} & \begin{tabular}[c]{@{}c@{}}Encode\\ Execution\\ Time\end{tabular} & \begin{tabular}[c]{@{}c@{}}Decode\\ Execution\\ Time\end{tabular} & \begin{tabular}[c]{@{}c@{}}Output\\ Domain\\ Names\end{tabular}                                                                                                                                                                                                                                                                                                                              \\ \hline
11                                                              & 24                                                              & 267.98s                                                           & 114.04s                                                           & \begin{tabular}[c]{@{}c@{}}lviazea01;\\ lviahgeurakk-04;\\ lvia79djqb02;\\ lviamhlayae2-;\\ lviabpi62f03;\\ lvielm13owul;\\ lvienh103;\\ lvwlei14j0volg002;\\ lvwltrti102\end{tabular}                                                                                                                                                                                                       \\ \hline
22                                                              & 44                                                              & 460.03s                                                           & 188.05s                                                           & \begin{tabular}[c]{@{}c@{}}lviaudzsud\\ zssic20e0002;\\ lvia523rajc10j02;\\ lvia76rptu16q603;\\ lviabpm01;\\ lvwleg186bczp;\\ lvienb-8702;\\ lvwlejo01;\\ lvwlt-28d6\\ gyopur24-05;\\ lvwlynaks;\\ lvwl07bsjm-\\ 1vusfoe-1ofw23\\ ucm29c58-05;\\ lvwlcko1-02\end{tabular}                                                                                                                    \\ \hline
34                                                              & 64                                                              & 781.06s                                                           & 191.32s                                                           & \begin{tabular}[c]{@{}c@{}}lviauket99jd104;\\ lviaha102;\\ lvia78h;\\ lviamiltathybpr5102;\\ lvienbpo-04;\\ lvwlep003;\\ lvwltb5eud01;\\ lvwl1d1601;\\ lvwlyunt9103;\\ lvwlxmanrovp;\\ lvwlcl67ds01;\\ lvw6l0v30ge-\\ nsigoyed6gy-01;\\ luctgold99q;\\ luctgjnbu9jv5wk;\\ luctiwaa;\\ luctigh5;\\ luctia-thne55x001;\\ luctin01;\\ luctivbusovalriq5ilf\\ 2l5lld-z2tchl0lcl2201\end{tabular} \\ \hline
\end{tabular}
\end{table}

\subsection{An illustrative example}
The HMM representing the real legitimate domain names is very complex (it has 2995 states and each state has at most 38 outgoing transitions). We use an example HMM (Figure \ref{ex-hmm}) to illustrate the algorithm. In the example HMM, the asymptotic probability of state $S_i$ ($i=0,1,2,3$) is calculated as: $P(S_0)=0.15$, $P(S_1)=0.46$, $P(S_2)=0.25$ and $P(S_3)=0.14$. So $S_1$ is chosen as the starting state for both the client and the server.

After the probability rounding algorithm, all transition probabilities are rounded to the closest $\frac{1}{2^n}$ (Figure \ref{round-fig}). After the binarization algorithm, each transition is associated with a binary representation (Figure \ref{bin-fig}).

Suppose that the `client message' is `h', its binary representation is `01101000'. The starting state is $S_1$. Following Figure \ref{bin-fig}, the state transition is $S_1$ $\rightarrow$ $S_1$ $\rightarrow$ $S_2$ $\rightarrow$ $S_3$ $\rightarrow$ $S_0$ $\rightarrow$ $S_1$ $\rightarrow$ $S_1$ $\rightarrow$ $S_1$. The mapping is in Figure \ref{map}. With the HMM-encoding algorithm, the input `h' is converted into `bdcabbb'.

\subsection{Performance}
We implemented the proof-of-concept idea on a laptop (OS: Windows 10, CPU: i5-5300U, RAM: 8G). We measure the encoding and decoding execution time on various lengths of input string. Table \ref{perf} shows the performance of the transparent protocol. For AES encryption/decryption, we use block size equal to 16. This will make the length of AESed-string 24 (when length of the input string is between 1-15), 44 (when length is between 16-31), 64 (when length is between 32 to 47) bytes etc. So the performance of the implementation is more dependent on the length of AESed-string than the length of the input string. Note that the input string can be any length.

From Table \ref{perf}, we can see that the execution time is proportional with the length of AESed-string. Note that, all output domain names start with `l', which is resulted from the same start state (the state with the largest asymptotic probability). How to pick different start states where the client and server agree on will be an interesting topic as the future work.

\begin{figure*}
\begin{center}
\includegraphics[width=0.9\textwidth]{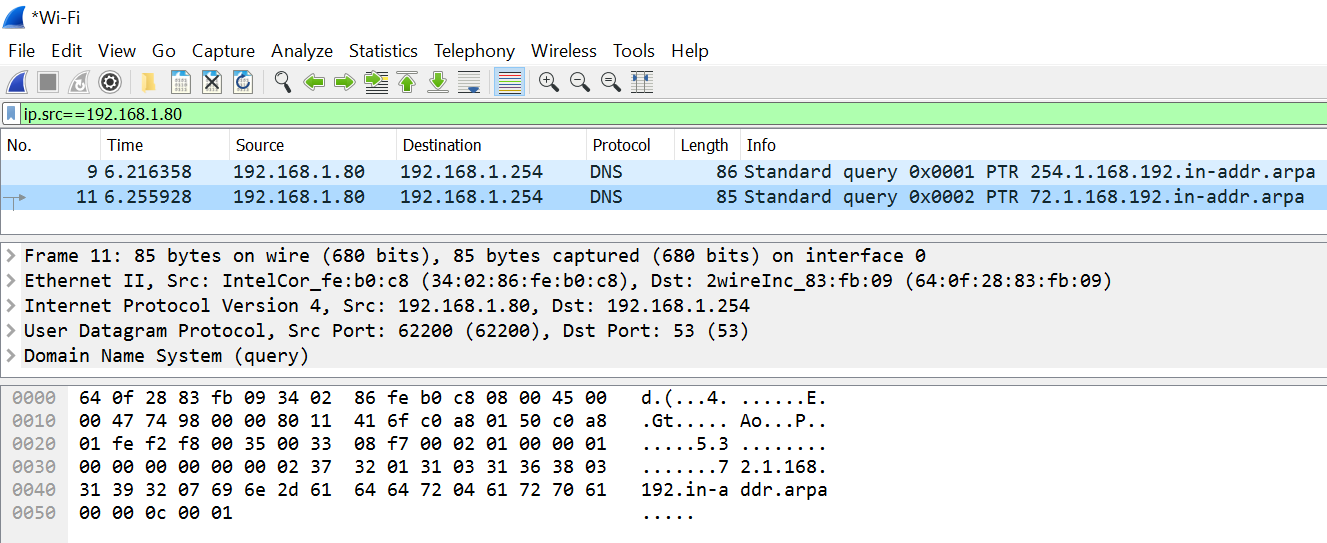}
\caption{Wireshark screenshot}
\label{wireshark}
\end{center}
\end{figure*}

Figure \ref{wireshark} shows the network traffic sniffed by wireshark. The server registers the domain names with a randomly chosen IP address (192.168.1.72). The client (192.168.1.80) is trying to retrieve a message from the server. All communication happens between the client to the local DNS server (192.168.1.254). And the DNS server cannot distinguish the DNS traffic from the others. This proves the destination IP address stays hidden and the idea works!

\subsection{Security Analysis}
To further validate the possibility of the implementation, we conduct security empirical analysis in terms of confidentiality, differentiability and communication.
\begin{itemize}[noitemsep,nolistsep]
\item Confidentiality: Since the arbitrary network traffic is masquerading as normal DNS traffic, it is hard to filter it out from other DNS traffic. Even if a man-in-the-middle (MITM) filters out the traffic and tried to decode the DNS packets, he/she has to have all pre-shared information: AES key, PRG, PRG seed, and the HMM. This is infeasible unless he has the software package.
\item Differentiability: Except for normal DNS queries, the application involves reverse-DNS lookup and DNS registration traffic. Reverse-DNS lookup is widely used by common security tools \cite{rdns} including network troubleshooting tools, anti-spam techniques, and system monitoring tools. For example, email anti-spam software checks the domain names using reverse-DNS lookup to see if the source is a dynamically assigned address, which is unlikely used by a legitimate mail server. Web browsers use reverse-DNS lookup to verify the same origin of the requests to avoid DNS rebinding attacks \cite{rdns2}. In terms of DNS registration traffic, although large companies like Google and Amazon keep registering domain names in a round-robin fashion, it remains unclear that how unusual the DNS registration traffic is. This will be an interesting topic for future work.
\item Communication: If there is an ISP-level surveillance tool looking for this application, it is impossible to pick up the related domain names. Even if it is possible, we can use Tor at the client side (who generates the domain names) to further hide the source IP address. Also, the registered IP address associated with the generated domain names is randomly selected, so there is no way to trace the communication parties using the IP address. 
\end{itemize}

\section{Conclusion and Future Work}\label{conclusion}
This paper proposes the structure of a new covert data transport protocol. It would be suitable for botnet C\&C. The advantages over existing protocol obfuscation tools are that network traffic is real and normal DNS lookup traffic on benign-looking domain names, which is protected against DPI detection. One possible disadvantage is the low throughput, which is more suitable for delay-tolerant communication than real-time chatting, such as twitter-like social networks, sensitive information retrieval tools etc. We hope this paper will provide a new insight into protocol obfuscation.

Future work can tune parameters to optimize the throughput including re-designing HMM encoding/decoding algorithm, better data structure to improve the performance etc.

The proposed concept will be helpful in moving botnet countermeasures to being pro-active. It will change the routines of botnets innovating, anti-virus vendors reverse-engineering and finding countermeasures \cite{fu2015analysis}. Instead, we need to develop better technology than the enemy and know how to counter botnets before they deploy innovations. Up to now, the botnet implementers have the advantage of creating the innovations. By designing countermeasures to better botnet designs in advance, it would make botnet deployment less profitable.

\section*{Acknowledgment}
This material is based in whole or in part upon work
supported by the National Science Foundation under Grant numbers ACI-1547164 and CNS-1544910. Any opinions, findings, and conclusions or recommendations expressed in this material are those of the author(s) and do not necessarily reflect the views of the National Science Foundation

{\small
\bibliographystyle{IEEEtran}
\bibliography{ref}}

\end{document}